\documentclass[aps,twocolumn,prd,nofootinbib,superscriptaddress,showkeys,english,final,balancelastpage]{revtex4}

\usepackage{graphicx}
\usepackage{epsfig,float}
\usepackage{appendix}
\usepackage{tabularx}
\usepackage{amssymb}
\usepackage{amsmath}
\usepackage{amsbsy}
\usepackage{comment}
\usepackage{mathrsfs}
\usepackage{dsfont}
\usepackage{nicefrac}
\usepackage{wrapfig}
\usepackage{amsfonts}
\usepackage{url} 
\usepackage[nolist]{acronym}
\usepackage{nicefrac}
\usepackage{subfigure}
\usepackage{epsf,color,colordvi,pifont}
\usepackage{showkeys}
\DeclareFontFamily{OT1}{pzc}{}
\DeclareFontShape{OT1}{pzc}{m}{it}{<-> s * [1.10] pzcmi7t}{}
\DeclareMathAlphabet{\mathpzc}{OT1}{pzc}{m}{it}

\def\deltabar{{\mathchar'26\mkern-9mu\delta}}

\begin{document}
\title{Magnetic-field-tunable  anisotropic blackbody radiation and condensation of slow thermal light  in dynamical  axion insulators }
\author{E. \surname{Kochems}}
\affiliation{Institut f\"{u}r Theoretische Physik, Heinrich-Heine-Universit\"{a}t D\"{u}sseldorf,\\ Universit\"{a}tsstra\ss e 1, 40225 D\"{u}sseldorf, Germany}
\author{G. \surname{Quintero Angulo}}
\affiliation{Institut f\"{u}r Theoretische Physik, Heinrich-Heine-Universit\"{a}t D\"{u}sseldorf,\\ Universit\"{a}tsstra\ss e  1, 40225 D\"{u}sseldorf, Germany}
\affiliation{Departamento de F\'isica Te\'orica, Facultad de F{\'i}sica, Universidad de la Habana, \\ San L{\'a}zaro y L, Vedado, La Habana 10400, Cuba}
\author{R. \surname{Egger}}
\author{C. \surname{M\"{u}ller}}
\author{S.  \surname{Villalba-Ch\'avez}}
\email{villalba@uni-duesseldorf.de}
\affiliation{Institut f\"{u}r Theoretische Physik, Heinrich-Heine-Universit\"{a}t D\"{u}sseldorf,\\ Universit\"{a}tsstra\ss e 1, 40225 D\"{u}sseldorf, Germany}

\begin{abstract}  

Thermal radiation features of dynamical axion insulators, which are characterized by an antiferromagnetic order with simultaneously broken time-reversal and space-inversion symmetries, are investigated. Planck's radiation law is shown to exhibit remarkable anisotropic behavior as a result of the strong dispersion caused by the light-matter interaction.  A crossover scenario at low temperature is identified and an associated phase highly populated by slow thermal photons  is revealed.   We show that the asymmetry degree of the heat radiation and its angular distribution can be controlled via a magnetic field, paving the way toward a directional-tunable mechanism for thermal quantum manipulation and storage.  Analogies  are drawn with the expected behavior of  blackbody radiation in the  core of neutron stars. 
\end{abstract}

\date{\today}

\maketitle

\section{Introduction}

Solving the strong $CP$ problem through the Peccei-Quinn mechanism led to the emergence of the QCD axion  \cite{Peccei:1977hh,Wilczek:1977pj,Weinberg:1977ma} and has since been a paradigm for the broader class of axions occurring in various standard model extensions  \cite{Dine:1981rt,zhitnitskii,kim,shifman,Meissner:2007xv,Witten:1984dg,Svrcek:2006yi,Lebedev:2009ag,LCicoli:2012sz}, some of which put them forward as viable candidates for nonbaryonic dark matter \cite{covi,Raffelt:2006rj,Duffy:2009ig,Sikivie:2009fv,Baer:2010wm}.  While the original QCD-axion was ruled out shortly after its prediction  \cite{Donnelly,Zehnder}, experimental endeavors toward the detection of axionlike particles are nowadays being carried out worldwide \cite{Irastorza,DiLuzio}, based predominantly on the  axion-diphoton  coupling that characterizes  the theoretical framework of axion electrodynamics (AED) \cite{Wilczek:1987pj}.  Despite compelling theoretical arguments supporting their existence, no axion has been detected to date, thus suggesting a feeble interplay between these elusive particles and the well-established standard model constituents.

With the plausible realization of dynamical axionlike fields in topological insulators with broken space-inversion and time-reversal symmetries, an enticing landscape to study  axion physics has been opened  \cite{Li}.  Indeed,  dynamical axion insulators (DAIs), such as the already synthesized  $\rm MnBi_2Te_4$  flakes \cite{Gong,Otrokov,CLiu,Klimovskikh,QiuNat}, constitute suitable platforms  for proof-of-principle tests of various phenomena on which axion searches rely. Moreover,  the intrinsic characteristics of the medium, such as stiffness and topology, along with the knowledge of the associated axion quasiparticle mass, make DAIs ideal for investigating phenomena beyond the perturbative regime that the counterpart of particle physics justifiably utilizes. Theoretical studies in this direction have revealed a few emerging phenomena closely linked to the hallmark medium's magneto-electric response \cite{Ooguri,Imaeda,Nomura,Taguchi,Sekine} and in line DAI-based methods to detect dark-matter axions have been proposed \cite{Marsh,Engel,QiuNat}. 

The thermal radiation properties of DAIs are of particular interest to nuclear astrophysics. This is due to their potential to offer significant insights into blackbody radiation within the magnetic dual chiral density wave (MDCDW) phase of dense quark matter \cite{Frolov, Efrain1,Efrain2,Efrain3,Efrain4}, which is presently regarded as a plausible model for characterizing the cores of neutron stars and magnetars \cite{EfrainUniverse,Gyory,Sadooghi}.  Axion electrodynamics rules light-matter interaction in the MDCDW  phase, just as it does in DAIs. As a result, the propagation of electromagnetic waves  in both scenarios is significantly influenced by  axion-polariton states \cite{EfrainNP,EfrainEPJC}. While some  astrophysical consequences of these hybridized  states  have been discussed, their significance for the stellar  thermal  properties remain uncertain.  However, the existing analogy between the MDCDW phase  and DAIs renders the latter an ideal playground for emulating hitherto unobserved phenomena of heat radiation under extreme conditions.

Moreover,  the  realization of cavity-based axion-polaritons \cite{Xiao} could benefit the quantum control of light-matter interaction over other widely studied systems,  like  qubits in superconducting circuits and cavity optomechanical oscillators \cite{Xiang,Aspelmeyer}.   The primary reasons for these prospects stem from DAI's seemingly favorable cooling feature and the opportunity to magnetically control the interaction strength \cite{Xiao2}.  However, high field strengths could induce noticeable  changes in the axion quasiparticle mass and thus modify the critical temperature for condensation of the massive branch of the axion-polariton ensemble, making it difficult to cool to the ground state even if it is characterized by a low population.  The question raised is whether further repercussions linked with the  breakdown of spatial isotropy may have nontrivial effects on the system's thermodynamic control. 

In this paper  we show  that,  near the critical boundary between the antiferromagnetic and paramagnetic phases,  the axion-polariton state renders  Planck's radiation law exhibiting a field-tunable anisotropy. This remarkable feature enables the control of direction-dependent macroscopic properties of thermal radiation under equilibrium conditions.   Our research uncovers qualitatively novel characteristics of DAIs, including a many-body condensate of slowly moving photons distinguished by low-dimensional transport of energy. Likewise, it is indicated, based on the aforementioned analogy, which of these characteristics prevail in the blackbody radiation of the MDCDW phase of dense quark matter.

The paper is structured in the following form.  In Sec.~\ref{sec2}, thermal field theory methods are employed to derive the partition function and Helmholtz free energy of the axion-photon ensemble interacting with a constant external magnetic field. The resulting formulas are subsequently utilized in Sec.~\ref{sec3} to obtain the modified Planck law, from which the nature of blackbody radiation in DAIs is elucidated. Our conclusions are given in Sec.~\ref{conclusions}. Details of the calculations and further information on the optical features of thermal radiation are provided in the  Appendix.

\section{Thermal Field Theory Approach \label{sec2}}

Let us consider a DAI  sample  characterized by a volume $V=L_xL_yL_z$.  We  describe the equilibrium blackbody radiation inside the insulator using AED.  By adopting rationalized  Gaussian  units with $c=\hbar=k_{\mathrm{B}}=\epsilon_0=1$, this  theoretical framework is characterized by the action  \cite{Li}
\begin{equation}
\begin{split}
 \Gamma&=\int d^4x\Big\{4\pi Jg^2\left[(\partial_t\delta\theta)^2-\left(v_i\partial_i\delta\theta\right)^2-m^2\delta\theta^2\right]\\
 &+\frac{1}{2}\left(\pmb{D}\cdot\pmb{E}-\pmb{H}\cdot\pmb{B}\right)+\frac{\alpha}{\pi}(\bar{\theta} +\delta\theta)\pmb{E}\cdot\pmb{B}\Big\}.
\end{split}\label{initial}
\end{equation}Here,  $\pmb{D}=\epsilon\pmb{E}$ and $\pmb{H}=\mu^{-1}\pmb{B}$ are the electric and magnetic displacement vectors, with $\epsilon$ and $\mu$ denoting the respective  dielectric constant and magnetic permeability, with  material stiffness $J$,  constant $g$, and  $\alpha=e^2/4\pi$ referring to  the fine structure constant. We note that the static axion field $\bar{\theta}$ takes $\bar{\theta}=\pi$  in time-reversal invariant topological insulators, whereas  $\bar{\theta}=0$  in topologically trivial insulators \cite{Sekine}.  In Eq.~\eqref{initial},  $v_{i=x,y,z}$  stands  for the group velocity components of the free low-energy spin wave $\delta\theta$  with  $\vert\delta\theta\vert<\bar{\theta}$, whereas  $m$ refers to its  mass.  This dynamical  field  is the projection of the fluctuation of the  spontaneous  antiferromagnetic order parameter, i.e., N\'eel's field, onto the crystallographic anisotropy axis of the magnetic topological insulator \cite{Engel}. In the following, we  employ a rescaled version of $\delta\theta$,  $\sqrt{8\pi Jg^2}\delta\theta\to \phi$,  which results in a multiplicative renormalization of the gauge coupling  $\alpha\to\alpha/\sqrt{8\pi Jg^2}$.  We remark that  Eq.~\eqref{initial} provides a reliable physical description when the energy and momentum of low-energy excitations are significantly below the material's bulk gap $\Delta\sim O(0.1)\;\rm eV$ and the upper momentum limit $\Lambda_i\sim\Delta/v_{\mathrm{i}}$, respectively. 

Heat radiation will be identified with small-amplitude electromagnetic waves $a_\mu(t,\pmb{x})$ describing the linear response of the system to the presence of a constant magnetic background $\pmb{B}=(0,0,B)$.  As a consequence, we will focus  on the action that results from Eq.~\eqref{initial}  when the electromagnetic fields $\pmb{E}$ and $\pmb{B}$ are replaced by  $\pmb{e}=-\pmb{\nabla}a_0-\partial \pmb{a}/\partial t$ and $\pmb{B}+\pmb{b}$ with $\pmb{b}=\pmb{\nabla}\times \pmb{a}$, respectively, and only  nontrivial bilinear combinations in $a_{\mu}(t,\pmb{x})$ are retained:
 \begin{equation}
\begin{split}
S&=\int d^4x\left\{-\frac{\pmb{B}^2}{2\mu}+\frac{1}{2}\left[\pmb{e}^2-c^{\prime 2}\pmb{b}^2\right]+\kappa\; \phi\; \pmb{e}\cdot\pmb{B} \right.\\
&\left.+\frac{1}{2}\left[(\partial_{t}\phi)^2-(v_i\partial_i \phi)^2-m^2\phi^2 \right]+\frac{\alpha\; \bar{\theta}(z)}{\pi\sqrt{\epsilon}}\; \pmb{e}\cdot\pmb{B} \right\},
\end{split}\label{bilinearaction}
\end{equation} where the rescaling $a_\mu(t,\pmb{x})\to a_\mu(t,\pmb{x})/\sqrt{\epsilon}$ has been carried out,  $c^\prime=1/\sqrt{\epsilon\mu}$ denotes  the speed of light in the medium, and  $\kappa=\alpha/\sqrt{(2\pi)^3Jg^2 \epsilon}$ is the coupling strength between $\phi(t,\pmb{x})$  and $a_\mu(t,\pmb{x})$ mediated by $\pmb{B}$.  Observe that, to account for the topological charges  on the surfaces  perpendicular to $\pmb{B}$, the  static axion field  $\bar{\theta}$  has been  promoted to a $z$-dependent function, i.e.,   $\bar{\theta}\to \bar{\theta}(z)=\bar{\theta}\,\Theta(L_z/2+z)\Theta(L_z/2-z)$, with $\Theta(x)$ denoting the unit-step function  [$\Theta(x)=1$ if $x\geqslant0$ and  $\Theta(x)=0$ otherwise].

Establishing the corresponding  Helmholtz free energy   $\mathscr{F}=-\beta^{-1} \ln \mathpzc{Z}$ requires adopting the imaginary time formalism  $t\to - i\tau$ with $0\leq \tau\leq \beta$ and $\beta=T^{-1}$ denoting the inverse temperature.  In this context,  the boson  fields $\phi(x)$  and $a_\mu(x)$ with $x_\mu=(\pmb{x},\tau)$ and $\mu=1,2,3,4$  are promoted to  periodic functions in the variable $\tau$.  
The calculation of the required partition function can be   carried out  by extending the covariant Faddeev-Popov ansatz \cite{Rivers,LeBellac} to the case under consideration: 
\[
\int \mathpzc{D}\chi \delta(\mathscr{G}(^\chi a_\mu])\mathrm{det}\left[ \delta \mathscr{G}[^\chi a_\mu]/\delta \chi\right)_{\chi=0}=1.
\]  Here  the gauge-transformed field is $^\chi a_\mu=a_\mu+\partial_\mu \chi$ and  $\mathpzc{D}\chi$  is the Haar's integration measure. The expression above combines 
 the gauge-fixing function   $\mathscr{G}[a_\mu]=-c^{\prime2}\pmb{\nabla}\cdot\pmb{a}-\partial_\tau a_4+\mathpzc{s}$ with $\mathpzc{s}(x)$ denoting an  arbitrary spacetime depending scalar. Moreover, $\mathrm{det}\left(\delta \mathscr{G}[^{\chi}a_\mu]/\delta \chi\right)_{\chi=0}= \mathrm{det}(-\square_{x,\tilde{x}})$  is  the  Fadeev-Popov determinant with   $\square_{x,\tilde{x}}\equiv\partial_x^2\delta^{4}(x-\tilde{x})$   and  $\partial^2= c^{\prime 2}\nabla^2+\partial_{\tau}^2$.   As such, the partition function is independent of $\mathpzc{s}(x)$.  Upon weighting it with the Gaussian factor  $\exp\left(-\frac{1}{2\zeta c^{\prime2}}\int_X d^4x\; \mathpzc{s}^2\right)$, with $\int_X d^4x\equiv \int_0^\beta d\tau\int_V d^3x$ and a gauge-fixing parameter  $\zeta$, and integrating functionally over  $\mathpzc{s}(x)$, the gauge-fixing term rises to the exponent, and  
\begin{equation}
\begin{split}\label{partitionfunction}
 \mathpzc{Z}&=\exp\left[-\frac{\beta V}{2\mu}B^2\right]\mathpzc{Z}_{\mathrm{AED}},\\
\mathpzc{Z}_{\mathrm{AED}}&=\mathrm{det}(-\square_{x,\tilde{x}})\int_{\mathrm{periodic}} \mathpzc{D}a \mathpzc{D}\phi\; e^{-S_{\mathrm{E}}}.
\end{split} 
\end{equation}Here $S_{\mathrm{E}}$ is the Euclidean action resulting from Eq.~\eqref{initial}  combined with the gauge-fixing term.

\begin{figure}
{\includegraphics[width=0.22\textwidth]{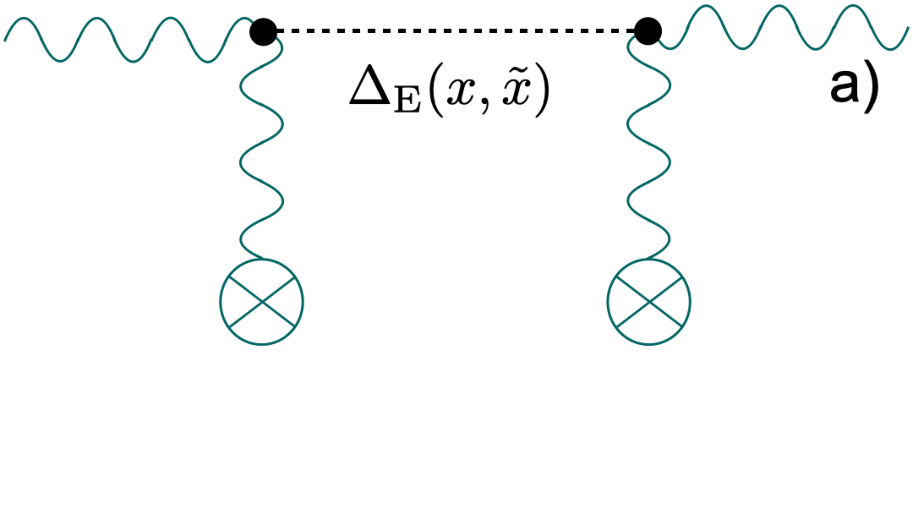}}
\includegraphics[width=0.22\textwidth]{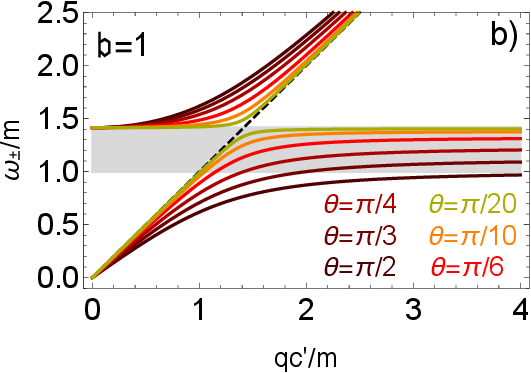}
\includegraphics[width=0.235\textwidth]{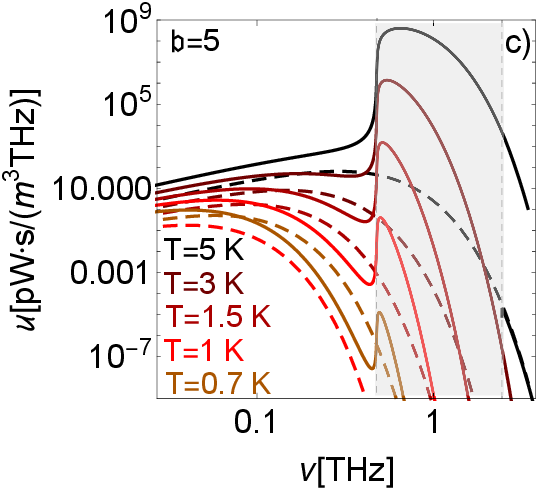}
\includegraphics[width=0.22\textwidth]{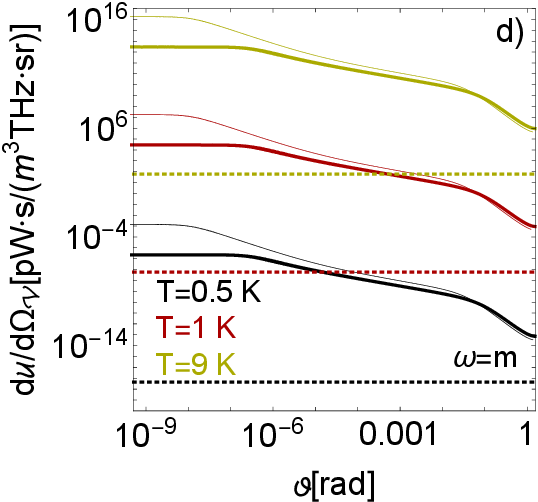}
\caption{(a) Polarization tensor in AED. (b) Dispersion relations. The dashed line is linked to the ordinary photon $\omega_{\mathrm{o}}$, whereas the gray band shows the maximum gap. While the lower (massless) branch is associated with extraordinary photons $\omega_{-}$,  the upper (massive) one is linked to axions $\omega_+$. (c) Blackbody spectrum. The gray band shows the maximum gap at $\mathfrak{b}=5$. The dashed (solid) curves display the ordinary (extraordinary) photon spectrum. The black solid curve in the lower right corner depicts the axion-like spectrum at  $T=5\;\mathrm{K}$. (d) Angular distribution of the internal energy density linked to extraordinary photons at $\omega=m$ [$d\Omega_{\mathpzc{v}}\equiv d\varphi d\cos(\vartheta)$]. Curves sharing a color are linked to a common temperature at $\mathfrak{b}=1$ (thick) and $\mathfrak{b}=5$ (thin). The dotted lines show the contributions of the ordinary mode.} 
\label{fig:1}
\end{figure}

The functional integrations in Eq.~\eqref{partitionfunction} can be straightforwardly calculated owing to the integrand's Gaussian nature. This feature guarantees, indeed, that the contribution of small amplitude waves dominates over the remaining ones. Upon carrying out the one over $\phi(x)$ and adopting the Feynman gauge $\zeta=1$, we find 
\begin{equation}
\begin{split}
&\mathpzc{Z}_{\mathrm{AED}}\propto\mathrm{det}(-\square_{x,\tilde{x}})[\mathrm{det}(-\square_{x,\tilde{x}}^v+m^2_{x,\tilde{x}})]^{-\nicefrac{1}{2}}\int \mathpzc{D}a\; \\
&\quad\times e^{-\int_X d^4x\left[\int_Xd^4\tilde{x}\;\frac{1}{2}a_\alpha(x)D^{-1}_{\alpha\beta}(x,\tilde{x})a_\beta(\tilde{x})+\varrho (z) a_4(x)\right]}.
\end{split}\label{Zeq}
\end{equation} Here   $\varrho(z)=\frac{\alpha B}{\sqrt{\epsilon}}[\delta(z+L_z/2)-\delta(z-L_z/2)]$ is the topological charge density, whereas  $D^{-1}_{\alpha\beta}(x,\tilde{x})=-\square_{x,\tilde{x}}\deltabar_{\alpha\beta}+\Pi_{\alpha\beta}(x,\tilde{x})$ denotes  the inverse photon propagator with  $\deltabar_{\alpha,\beta}=\mathrm{diag}(1,1,1,1/c^{\prime 2})$ and  \[
\Pi_{\alpha\beta}(x,\tilde{x})=-\kappa^2\tilde{F}_{\alpha\sigma}\left[\partial_\sigma^{x}\partial_\mu^{\tilde{x}}\Delta_{\mathrm{E}}(x,\tilde{x})\right]\tilde{F}_{\mu\beta}
\]  the polarization tensor [see Fig.~\ref{fig:1}(a)] mediated by the Euclidean axion propagator $\Delta_{\mathrm{E}}(x,\tilde{x})$ \cite{Selym1,Selym2}.  Here, $\tilde{F}_{\alpha\beta}=\frac{1}{2}\varepsilon_{\alpha\beta\mu\nu}F_{\mu\nu}$ with $\varepsilon_{1234}= 1$. The components of  the external electromagnetic tensor  $F_{\alpha\beta}$ are $F_{j4}=0$ and $F_{j\ell}=-\epsilon_{j\ell k} B_k$.  In Eq.~\eqref{Zeq} we have used the shorthand notation  $\square_{x,\tilde{x}}^v\equiv [\partial_\tau^2+(v_i\partial_i)^2]\delta^{4}(x-\tilde{x})$  and  $m_{x,\tilde{x}}^2\equiv m^2\delta^{4}(x-\tilde{x})$.  Next  we express $a_\alpha(x)$ as a Fourier series  $a_\alpha(x)=\sqrt{\beta/V}\sum_{n,\pmb{q}}a_{\alpha}(n,\pmb{q})\exp[i(\omega_\mathrm{n}\tau+\pmb{q}\cdot\pmb{x})]$ where $\omega_\mathrm{n}=2n\pi/\beta$  are  the bosonic  Matsubara's frequencies,  
 and  carry out  the  integration.  As a consequence, 
\begin{equation}
\begin{split}
\mathpzc{Z}_{\mathrm{AED}}&=\exp\left[-\varsigma\frac{\alpha^2}{\pi^3\epsilon}\bar{\theta}^2\beta V  B^2\right]\\
&\times\prod_{\mathrm{n},\pmb{q}}[\beta^2(\omega_\mathrm{n}^2+\omega_{\mathrm{o}}^2)][\beta^2 (\omega_{\mathrm{n}}^2+v_i^2q_i^2+m^2)]^{-\nicefrac{1}{2}}
\\&\times\{\beta^8\mathrm{det}[\deltabar_{\alpha\beta}(\omega_\mathrm{n}^2+\omega_{\mathrm{o}}^2)+\Pi_{\alpha\beta}(-q)]\}^{-\nicefrac{1}{2}},
\end{split}\label{finalZaed}
\end{equation}where $\varsigma=1-\mathrm{Si}(1)+\pi/2-\cos(1)\approx1.084410951$,  with $\mathrm{Si}(x)$ referring to the sine integral function.  Here    $\omega_{\mathrm{o}}=c^\prime\vert\pmb{q}\vert$ is the ordinary photon dispersion law in the medium. In Eq.~\eqref{finalZaed} an unessential  proportionality factor has been ignored   and the Fourier transform of the  polarization tensor reads $\Pi_{\alpha\beta}(\tilde{q},q)=\delta_{\tilde{q},-q}\mathpzc{P}_{\alpha\beta}(q)$, where 
 $\Pi_{\alpha\beta}(q)=\kappa^2\Delta_{\mathrm{E}}(q)\tilde{F}_{\alpha\lambda}q_\lambda \tilde{F}_{\beta\sigma} q_\sigma$ with 
$ \Delta_{\mathrm{E}}(q)=(\omega_{\mathrm{n}}^2+v_i^2q_i^2+m^2)^{-1}$  and $q_\mu=(\pmb{q},\omega_{\mathrm{n}})$.  

Next we substitute  Eq.~\eqref{finalZaed} into Eq.~\eqref{partitionfunction} and obtain, from the resulting expression for $\mathpzc{Z}$, the  Helmholtz free energy. After carrying out the sum over the Matsubara  frequencies we find 
\begin{equation}\label{freeenergy}
\mathscr{F}=\frac{V}{2\mu}B^2\left[1+2\varsigma\frac{\alpha^2}{\pi^3}\frac{\mu}{\epsilon}\bar{\theta}^2\right]+\mathscr{F}_{\mathrm{vac}}+\mathscr{F}_{\mathrm{st}}
\end{equation} with  $\mathscr{F}_{\mathrm{vac}}=\frac{1}{2}V\sum_{i}\int_{\pmb{\Lambda}} \frac{d^3q}{(2\pi)^3}\omega_{i}$ the vacuum contribution and  $\mathscr{F}_{\mathrm{st}}=\frac{V}{\beta}\sum_{i}\int_{\pmb{\Lambda}} \frac{d^3q}{(2\pi)^3}\ln\left(1-e^{-\beta\omega_{i}}\right)$ the  statistical one.  Here we  take the continuum limit $V\to\infty$, bounding the integration domain to a region $\pmb{\Lambda}=(\Delta/v_x,\Delta/v_y,\Delta/v_z)$,  where the low-energy description \eqref{initial} applies. In $\mathscr{F}_{\mathrm{vac}}$ and $\mathscr{F}_{\mathrm{st}}$,  the latin index $i$   runs over  $i=\mathrm{o},\,\pm$, covering the massless (photonlike excitations) and massive branches of the polariton state, which are linked  with the extraordinary  ($\omega_{-}$) and axionlike ($\omega_{+}$) dispersion relations  
 \begin{equation}
 \omega_{\mp}^2 =\frac{1}{2}(\omega_{\mathrm{o}}^2+w_{*}^2)\mp\frac{1}{2}\sqrt{(\omega_{\mathrm{o}}^2-w_{*}^2)^2+4m^2\mathfrak{b}^2c^{\prime 2}q_\perp^2},\label{disperisonrelationpolariton}
 \end{equation} respectively. Here $q_\perp$ is  the  momentum   perpendicular to $\pmb{B}$   
 and  $w_{*}=(v_i^2q_i^2+m_*^2)^{1/2}$ with  $m_*=m(1+\mathfrak{b}^2)^{1/2}$   the  dressed axion mass,   $\mathfrak{b}=B/B_{0}$ and   $B_{0}=m/\kappa$ denoting the characteristic magnetic field scale of AED. 

\section{Results and Discussion \label{sec3}}

Our numerical evaluation relies on the benchmark parameters estimated for  the antiferromagnetic phase of  $(\mathrm{Bi}_{1-x}\mathrm{Fe}_x)_2\mathrm{Se}_3$ with a nominal doping concentration approximately equal to $3.5 \%$ \cite{Li,Zhang}. In this case, the factor $Jg^2\approx 450\; \rm eV^2$ and the dynamical axion mass can reach values as small as  $m\approx2\;\rm meV$ near the critical boundary between the antiferromagnetic and paramagnetic phases. The values of these parameters are generally $T$-dependent  \cite{WanSL,Ishiwata1}, lifting the typical mass of the axion-like quasiparticle  to   $m\sim O(1)\; \rm eV$  far from the phase boundary \cite{Ishiwata1,Ishiwata2}. Conversely, in the vicinity of the critical limit, the mass changes with the doping concentration but varies very slowly with temperature, enabling us to treat the benchmark mass as a constant \cite{WanSL}.   Moreover, the  estimated dielectric constants in this material  are  $\epsilon\approx 25$ and  $\mu\sim1$, leading to a  speed of light   $c^\prime\approx 0.2$ and a characteristic magnetic field  scale $B_{0}\approx 2.35\; \rm T$ that can easily  be surpassed.  Hereafter, we will consider the velocity components of a  free  axion  to be equal and of the order of the spin wave speed, which we take  as $v_{x,y,z}=v_{\mathrm{s}}= 10^{-4}$ \cite{Kanj,Engel,Pickart}.   Figure~\ref{fig:1}(b) shows the  mutual repelling behavior of the resulting dispersion relations \label{red}{[see Eq.~\eqref{disperisonrelationpolariton}]} that characterize the axion-polariton state.   Here  $\theta\in[0,\pi]$ is the angle between the wave vector $\pmb{q}$  and $\pmb{B}$.

 At this point, we find it convenient to determine the inverse relations  $\vert \pmb{q}\vert_\pm=\mathpzc{n}_\pm \omega$ with $\mathpzc{n}_\pm$ referring to  the  refraction indices, explicitly, 
\begin{equation}\label{refraction_indices}
\begin{split}
&\mathpzc{n}_\pm^2(\omega,\theta)=\frac{1}{2v_{\mathrm{s}}^2}\left[1+\frac{v_{\mathrm{s}}^2}{c^{\prime2}}-\frac{m_*^2(\theta)}{\omega^2}\right]\\
&\quad\mp\frac{1}{2v_{\mathrm{s}}^2}\sqrt{\left[1+\frac{v_{\mathrm{s}}^2}{c^{\prime2}}-\frac{m_*^2(\theta)}{\omega^2}\right]^2-4\frac{v_{\mathrm{s}}^2}{c^{\prime2}}\left[1-\frac{m_*^2}{\omega^2}\right]},
\end{split}
\end{equation} where $m_*^2(\theta)\equiv m^2[1+\mathfrak{b}^2\cos^2(\theta)]$ and as  before the negative (positive) subindex is linked to the massless (massive) branch because $\lim_{\omega\to0}\vert\pmb{q}\vert_-=0$, whereas $\lim_{\omega\to m_*}\vert\pmb{q}\vert_+=0$.  The expressions for $\vert \pmb{q}\vert_\pm$ as well as $\omega_\pm$ in Eq.~\eqref{disperisonrelationpolariton} are solutions of the dispersion equation
\begin{equation}
\begin{split}\label{DE}
&\omega^4-\omega^2\left(\omega_o^2+w_*^2\right)+ \omega_o^2w_*^2 \\
&\qquad\qquad\qquad\qquad\qquad-m^2\mathfrak{b}^2c^{\prime 2}\vert\pmb{q}\vert^2\sin^2(\theta)=0,
\end{split}
\end{equation}from which  the relevant group-velocity components are obtained.   The ones  perpendicular and parallel   to $\pmb{B}$ read
\begin{equation}\label{velocities_components}
\begin{split}
&\mathpzc{v}_{\perp,i}(\omega,\theta)=\left\vert\frac{\partial \omega_i}{\partial q_\perp}\right\vert_{\vert\pmb{q}\vert\to\mathpzc{n}_i\omega}=c^{\prime2}\mathpzc{n}_{i}\sin(\theta)\\
&\qquad\qquad\times\frac{\omega^2\left[1+\frac{v_{\mathrm{s}}^2}{c^{\prime2}}-2 v_{\mathrm{s}}^2\mathpzc{n}_i^2\right]-m^2}{2\omega^2\left[1-\frac{1}{2}\left(1+\frac{v_{\mathrm{s}}^2}{c^{\prime2}}\right)\frac{\mathpzc{n}_i^2}{\mathpzc{n}_o^2}\right]-m_*^2},\\
& \mathpzc{v}_{\parallel,i}(\omega,\theta)=\left.\frac{\partial \omega_i}{\partial q_\parallel}\right\vert_{\vert\pmb{q}\vert\to\mathpzc{n}_i\omega}=c^{\prime2}\mathpzc{n}_{i}\cos(\theta)\\
&\qquad\qquad\times\frac{\omega^2\left[1+\frac{v_{\mathrm{s}}^2}{c^{\prime2}}-2 v_{\mathrm{s}}^2\mathpzc{n}_i^2\right]-m_*^2}{2\omega^2\left[1-\frac{1}{2}\left(1+\frac{v_{\mathrm{s}}^2}{c^{\prime2}}\right)\frac{\mathpzc{n}_i^2}{\mathpzc{n}_o^2}\right]-m_*^2},
\end{split}
\end{equation}
where the replacement $\vert\pmb{q}\vert\to\mathpzc{n}_i\omega$ must be carried out only after the differentiation  and $i=\pm$.   The derivatives in Eq.~\eqref{velocities_components} can easily be  established by implicitly differentiating  Eq.~\eqref{DE}  with respect to the variables  $q_{\perp,\parallel}$. 

\subsection{Anisotropic blackbody spectra}

In order to assess how  the  field-dependent dispersion phenomenon affects the medium's thermal properties, we first focus on the blackbody spectra.  To  this end, we primarily investigate the internal energy density 
\begin{equation}\label{IEnergy}
\begin{split}
\mathpzc{U}&=\frac{1}{V}\frac{\partial(\beta \mathscr{F})}{\partial\beta}=\frac{B^2}{2\mu}\left(1+2\varsigma\frac{\alpha^2}{\pi^3}\frac{\mu}{\epsilon}\bar{\theta}^2\right)+\sum_i\mathpzc{U}_i,\\
\mathpzc{U}_i&=\int_{\pmb{\Lambda}} \frac{d^3q}{(2\pi)^3}\frac{\omega_i}{\exp(\beta\omega_i)-1}+\frac{1}{2}\int_{\pmb{\Lambda}}  \frac{d^3q}{(2\pi)^3} \omega_i.
\end{split}
\end{equation}By  going over to spherical variables and neglecting the vacuum contribution, 
\begin{equation}\label{MPRL}
\begin{split}
u_i&=\frac{d\mathpzc{U}_i}{d\nu}=\frac{4\pi^2}{c^{\prime2}}\frac{\nu^3}{e^{2\pi\nu/T}-1}\int_0^\pi d\theta \frac{\mathpzc{n}_{i}^2}{\mathpzc{n}_{o}^2} \frac{\sin(\theta)}{\vert \mathpzc{v}_{\pmb{q},i}\vert}.
\end{split}
\end{equation}Here $\mathpzc{n}_i=\vert\pmb{q}\vert/\omega_i$ and $\pmb{\mathpzc{v}}_i=\pmb{\nabla}_{\pmb{q}}\omega_i$ stand for mode $i$'s refraction index and group velocity, respectively.  The  factor  $\vert\mathpzc{v}_{\pmb{q},i}\vert^{-1}=\vert\pmb{\mathpzc{v}}_{i}\cdot\hat{\pmb{n}}\vert^{-1}$  is the Jacobian resulting from  adopting  $\nu=\omega/2\pi$  as an  integration variable. The expression above resembles  Planck's radiation law for dispersive anisotropic media  \cite{Cole,Mercier}.   Figure~\ref{fig:1}(c) shows the  behavior of  $u_{i}$ in a DAI when   $\mathfrak{b}=5$.     The  frequency range covers quasiparticle energies $\omega<10\;\rm meV$   below  the material's bulk gap approximately $O(0.1)\;\rm eV$ as required by the   low-energy description.  Moreover, the sample extension  $L_{x,y,z}\sim O(1)\;\rm  cm$  limits  $\vert q_{x,y,z}\vert\gg O(10)\;\rm \mu eV$, providing the lowest frequency bound  $\nu_L \gg O(10^{-1})\;\rm GHz$, corresponding to $\omega\gg O(1)\;\rm \mu eV$ \cite{Li,Engel},  from which the results are expected to be reliable.  As one could anticipate, the extraordinary spectrum shows a remarkable departure  as  $\nu\to m/2\pi\approx 0.48\;\rm THz$ from the left,  exhibiting a spectral  peak whose height increases with temperature. These findings suggest that there is a characteristic temperature at which the latter becomes stronger than Wien's maximum, leading to a phase transition of the system.   We will soon return to this point.

We point out that  the characteristic  flattening in the dispersion relation of  the extraordinary mode   implies that  corresponding quanta with $q_\perp>m/c^\prime$ are characterized by a perpendicular group velocity component $\mathpzc{v}_{\perp,-}=\vert\partial\omega_-/\partial q_\perp\vert\ll c^\prime$. Thus, in contrast to ordinary photons,  the direction of energy transport of the extraordinary and axion-like modes differs  from their respective wave vectors. If $\vartheta=\tan^{-1}(\mathpzc{v}_{\perp,-}/\mathpzc{v}_{\parallel,-})$ and $\theta=\tan^{-1}(q_\perp/q_\parallel)$ are the respective  group and phase velocity angles  relative to  $\pmb{B}$ with $\mathpzc{v}_{\parallel,-}=\partial\omega_-/\partial q_\parallel$,  then 
\begin{equation}
\tan(\vartheta)=\frac{\mathpzc{v}_\perp}{\mathpzc{v}_\parallel}=\tan(\theta)\frac{\omega^2\left[1+\frac{v_{\mathrm{s}}^2}{c^{\prime2}}-2 v_{\mathrm{s}}^2\mathpzc{n}_-^2\right]-m^2}{\omega^2\left[1+\frac{v_{\mathrm{s}}^2}{c^{\prime2}}-2 v_{\mathrm{s}}^2\mathpzc{n}_-^2\right]-m_*^2},\label{relatiorn}
\end{equation}where the results in Eq.~\eqref{velocities_components} have been used. The established connection enables us to parametrically generate a graph  exhibiting how the energy density due to extraordinary photons  varies with the angle $\vartheta$ (solid curves in  Fig.~\ref{fig:1}(d).     Observe that the solid curves tend to be  peaked at $\vartheta\approx0$  and that this behavior is more prominent the stronger the field is.   This implies that,  even though the states of extraordinary quanta might be characterized by a non-zero $q_\perp$ with  $m/c^{\prime}<q_\perp<O(0.1)\;\rm eV$, the energy distribution occurs asymmetrically, tending to gather extraordinary photons along the  $\pmb{B}$ axis. A similar effect is known in QED and has since proven crucial for understanding the photon capture effect and its role within the emission mechanism of pulsars \cite{shabadnat,shabad3v,shabad2004}.

\begin{figure}
\includegraphics[width=0.23\textwidth]{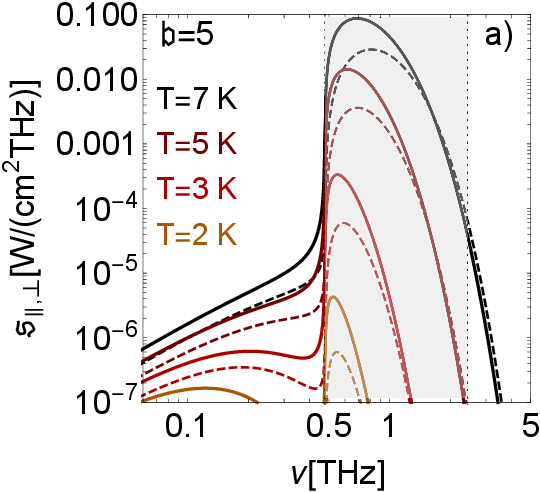}
 \includegraphics[width=0.23\textwidth]{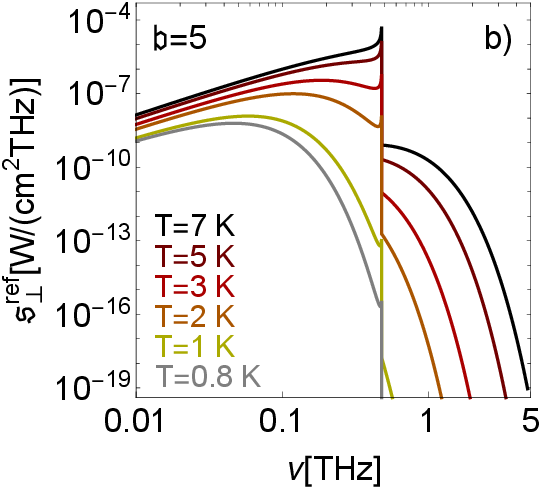}
 \includegraphics[width=0.23\textwidth]{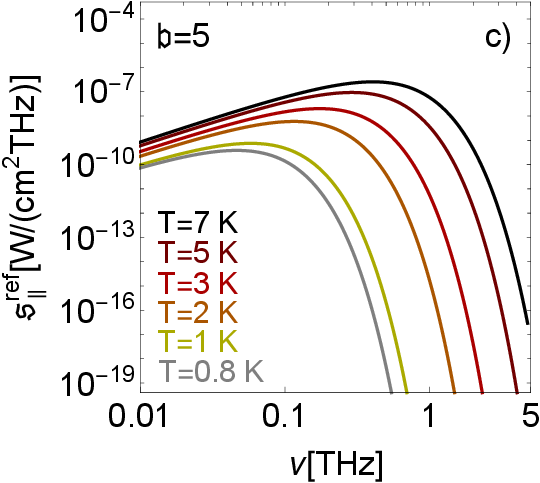}
 \includegraphics[width=0.225\textwidth]{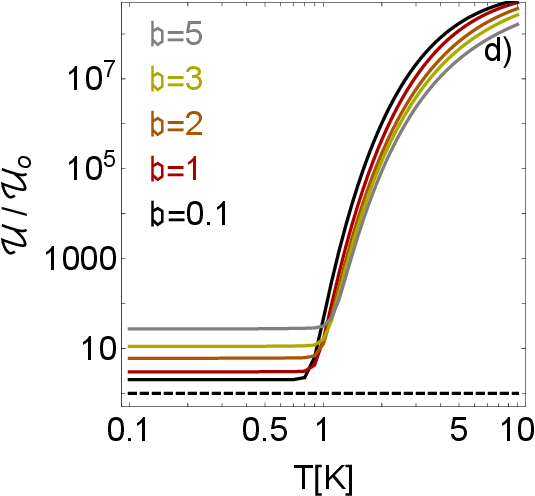}
\caption{ (a) Spectral intensity distributions parallel (solid) and perpendicular (dashed) to $\pmb{B}$ at $\mathfrak{b}=5$ for different temperatures. The gray band shows the maximum frequency gap. Also shows is the spectral intensity  (b) perpendicular and (c) parallel to  the magnetic-field direction, which can undergo refraction at the DAI-vacuum interface. (d) Dependence of internal energy density $\mathpzc{U}$ on the system's temperature for various magnetic fields. The  dashed line gives for comparison a ratio of unity.} 
\label{fig:2}
\end{figure}

To assess the consequence of  this effect, we have examined the spectral fluxes $\mathfrak{s}_{\parallel,\perp}=\frac{1}{V}\frac{dI_{\parallel,\perp}}{d\nu}$   parallel and perpendicular to  $\pmb{B}$ with $ I_{\parallel,\perp}$ denoting the corresponding radiation intensities. The previous quantities are linked to  the Cartesian components of the microscopic Poynting vectors $(\pmb{s}_i=\omega_i\pmb{\mathpzc{v}}_i)$ of each propagating  mode, and  indeed, decompose $\mathfrak{s}_{\parallel,\perp}=\sum_i\mathpzc{s}_{\parallel,\perp, i}$ with   
\begin{equation}
\mathpzc{s}_{\parallel,\perp, i}=\frac{2\pi}{c^{\prime 2}}\frac{\nu^3}{e^{2\pi\nu/T}-1}\int_{\Sigma_{\parallel,\perp}}d\Omega\;\frac{\mathpzc{n}_i^2}{\mathpzc{n}_o^2}\frac{\mathpzc{v}_{\parallel,\perp,i}}{\vert \mathpzc{v}_{\pmb{q},i}\vert}.
\end{equation} Here  $d\Omega\equiv d\varphi d\cos(\theta)$ is an element of the solid angle and $\Sigma_{\parallel,\perp}$ are the corresponding integration regions, chosen so that the  group-velocity components are positive  (see the Appendix). The $T$ dependence of  $\mathfrak{s}_{\parallel}$ (solid curves) and $\mathfrak{s}_\perp\equiv \mathfrak{s}_x=\mathfrak{s}_y$ (dashed curves)  is shown in Fig.~\ref{fig:2}(a)  for $\mathfrak{b}=5$. For a particular temperature, $\mathfrak{s}_\parallel$ and $\mathfrak{s}_\perp$ are qualitatively comparable but quantitatively  different.  The behaviors shown in Fig.~\ref{fig:2}(a) must be distinguished from the corresponding portions that may refract at the DAI-vacuum interfaces and, as a result, be detected [see Figs.~\ref{fig:2}(b) and \ref{fig:2}(c)].  The critical angles for internal reflections related to the ordinary and extraordinary photons, both parallel and perpendicular to the magnetic field, play important roles in achieving these results (see details in  the Appendix). In contrast to those associated with ordinary radiation, the ones linked to extraordinary thermal photons are dependent on $\omega$. For  $\nu<0.48\;\rm THz$  ($\omega<m$),  no angular restrictions occur on $\mathfrak{s}_\perp^{\mathrm{ref}}$, and all extraordinary quanta can undergo refraction at the DAI-vacuum interface, which is perpendicular to the  $x$ axis. The fact that the corresponding sector in Fig.~\ref{fig:2}(b)   matches that of  Fig.~\ref{fig:2}(a)  indicates that in the infrared regime, the vast majority of the refracted radiation through the latter plane is of extraordinary origin. The situation is different when $\nu\geqslant  0.48\;\rm THz$  $(\omega \geqslant m)$. In this case, $\mathfrak{s}_\perp^{\mathrm{ref}}$ is dominated by  ordinary photons  because the integration over $\theta$ in the extraordinary contribution is constrained to an infinitesimal region. This justifies the abrupt change exhibited in Fig.~\ref{fig:2}(a) close to  $\nu=0.48\; \rm THz$.

 Unlike Fig.~\ref{fig:2}(b),   Fig.~\ref{fig:2}(c) shows that the behavior of   $\mathfrak{s}_\parallel^{\mathrm{ref}}$  resembles the classical blackbody shape. A direct comparison with the result presented in the right sector  ($\nu\geqslant  0.48\;\rm THz$) of Fig.~2 indicates that the spectral intensity parallel $\pmb{B}$, which may undergo refraction at the DAI-vacuum interface,  predominantly originates from  extraordinary quanta. That $\mathfrak{s}_\perp^{\mathrm{ref}}$ and $\mathfrak{s}_\parallel^{\mathrm{ref}}$ differ from each other indicates that the emission of thermal radiation from the material is also anisotropic. Thus,  photon detection parallel and perpendicular to  $\pmb{B}$ will occur at different rates, providing indirect confirmation of the internal DAI features.   Measurements might be  conducted  using  microbolometers based on superconducting nanowires, which are projected to energy-resolve single-photon detection as low as $0.12\;\rm THz$ \cite{Wei,Karasik}. At $T=7\; \rm K$, $\mathfrak{b}=5$,  $\omega$ slightly  below  $2\;\rm meV$, and assuming a detection area $A\sim \mu\rm m^2$, we estimate  maximum output powers of  $P_\parallel=I_\parallel A\sim 2\times 10^{-15}\;\rm W$ and  $P_\perp=I_\perp A\approx 10 P_\parallel$, corresponding to  upper bounds of $6$ and  $60$ emitted photons per microsecond, respectively.  
  
We remark that the fact that  $\mathfrak{s}_\perp^{\mathrm{ref}}$  maximizes near $\omega\sim m$ for  certain temperatures,  accomplishing the accepted criterion  $T\ll\omega$  for high sensitivity levels in DAI-based searches for axion dark matter \cite{Marsh,Engel}, highlights the significance of blackbody radiation for the planned experimental setups for $T\sim O(1)\;\rm K$. Indeed, our analysis predicts that---for such temperatures---a background of thermal photons with frequency near  the detection frequency $\omega \sim m_*$ could affect their discovery potential based on photon counting. This difficulty can be  minimized by performing measurements at much lower temperatures $T\sim O(10^{-2})\; \rm K$ where single-photon detection in the terahertz regime is still possible \cite{Komiyama}.

\subsection{Condensation of slow thermal light}

\begin{figure}
\includegraphics[width=0.22\textwidth]{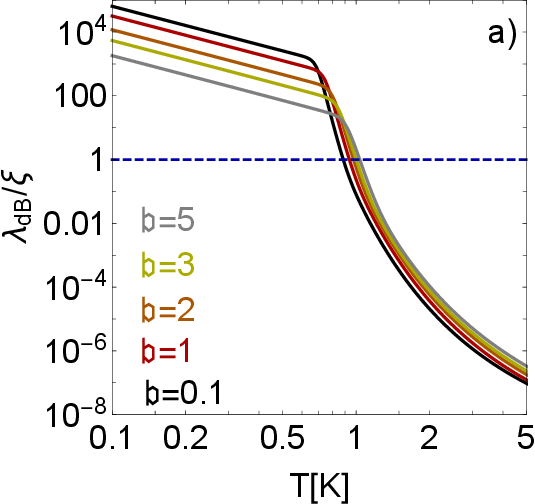}
\includegraphics[width=0.22\textwidth]{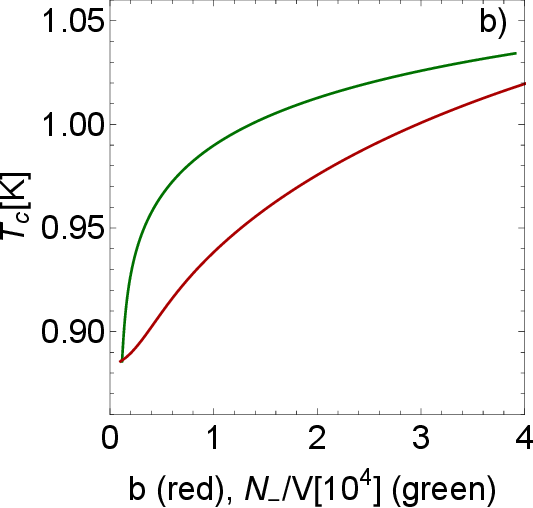}
\includegraphics[width=0.22\textwidth]{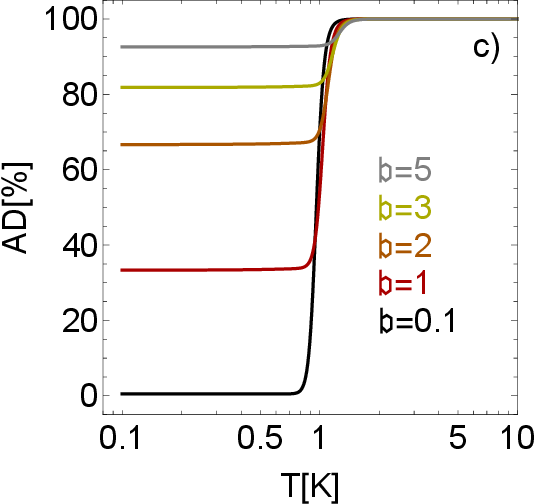}
\includegraphics[width=0.235\textwidth]{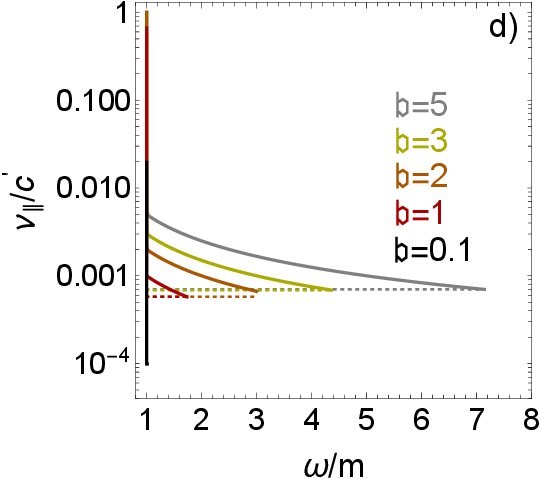}
\caption{ (a) Dependence of the ratio between the thermal de Broglie wavelength and the interparticle distance with temperature for various field strengths.  (b) Critical temperature as a function of the magnetic field parameter (red) and the density of extraordinary photons (green).   (c) Asymmetry  degree of heat radiation in DAI vs temperature for different $\mathfrak{b}$. (d) Behavior of the group velocity component of extraordinary photons along the magnetic field for $\omega \geqslant \gamma m$.} 
\label{fig:3}
\end{figure}

As the area below the extraordinary spectrum [see Fig.~\ref{fig:1}(c)]  increases   with temperature, its contribution to the  energy density  outweighs those of  the remaining propagation modes. Figure~\ref{fig:2}(d) exhibits the $T$ dependence of the total energy density in units of  $\mathpzc{U}_o=\frac{8\pi^2}{c^{\prime3}}\int_{\nu_{L}}^{\nu_\Lambda}\frac{ d\nu\;\nu^3}{\exp(2\pi\nu/T)-1}$  for magnetic-field strengths that fall into the range  permitted in the nontrivial topological phase  \cite{Xiao}.  To meet the limitations outlined below Eq.~\eqref{MPRL}, the lowest integration  limit  here and in the contribution of  the extraordinary mode   was taken as  $\nu_{L}=10\;\rm \rm GHz$,   whereas $\nu_\Lambda=10\;\rm meV/2\pi$ follows from the  highest energy cutoff. 
The findings  indicate  a crossover  at  $0.89\;\mathrm{K}\leqslant T_{\mathrm{c}}\leqslant 1.03\;  \rm K$  between regions with different  phenomenologies.  The existence of a critical temperature is confirmed by matching the thermal de Broglie wavelength  $\lambda_{\mathrm{dB}}=2\pi/\langle\vert\pmb{q}\vert\rangle$ and the characteristic distance between the extraordinary photons $\xi\approx (V/N_{-})^{1/3}$ [see Fig.~\ref{fig:3}(a)].   The intersections of the horizontal dashed line and the solid curves confirm  the existence of critical temperatures.  Deriving an approximate expression for $T_{\mathrm{c}}$ is a challenging endeavor due to the complicated nature of the dispersion relation and its $B$ dependence.  However, a numerical evaluation is shown in  Fig.~\ref{fig:3}(b), which  reveals a nonlinear dependence on $N_-$ and $B$. 

Returning  to Fig.~\ref{fig:2}(d), for $\lambda_{\mathrm{dB}}>\xi$, corresponding to  $T<T_{\mathrm{c}}$, $\mathpzc{U}_{-}+\mathpzc{U}_{+}\approx \mathpzc{U}_{-}\approx\mathcal{C}\mathpzc{U}_o$,   where $\mathcal{C}$ is a $B$-dependent constant because, at lower temperatures, the area below the extraordinary spectrum [see Fig.~\ref{fig:1}(c)]  is still dominated by a Planck profile, although dressed by the   factor $\mathpzc{n}_{-}^2/\vert\mathpzc{v}_{\pmb{q},-}\vert$.  The $T$ scaling of $\mathpzc{U}_{-}$ for  $T>T_{\mathrm{c}}$, corresponding to $\lambda_{\mathrm{dB}}<\xi$,  is  however  significantly stronger than $\mathpzc{U}_o\approx\pi^2 T^4/(30 c^{\prime3})$, outweighing the contribution of the ordinary photon gas by  various   orders of magnitude.   This  indicates that the number of extraordinary photons exceeds the ordinary ones.  This statement is supported  by Fig.~\ref{fig:3}(c) which shows the asymmetry degree (AD) equal to $(N_{-}-N_o)/(N_-+N_o)$, with  
 \[
N_{i}=\frac{2\pi V}{c^{\prime 2}}\int_{\nu_L}^{\nu_\Lambda}\frac{d\nu\; \nu^2}{\exp(2\pi\nu/T)-1}\int_0^\pi \frac{d\theta\,\sin(\theta)}{\vert\mathpzc{v}_{\pmb{q},i}\vert}\frac{\mathpzc{n}_i^2}{\mathpzc{n}_o^2}
\] defining  the mean  $i$-particle number. The AD  can be controlled  efficiently by  the magnetic field when  $T<T_{\mathrm{c}}$,  whereas  for $T>T_{\mathrm{c}}$  an almost  complete asymmetry approximately equal to  $100\%$  is reached, regardless of $B$.   Thus, for $\mathfrak{b}<1$,  the subcritical regime  $T<T_{\mathrm{c}}$ is a disordered phase because the photon species are in balance and have a broad spectrum. The latter feature persists for  $\mathfrak{b}>1$ and $T<T_{\mathrm{c}}$, but the phase  becomes asymmetric due to  the strong dispersion. In contrast,  an  ordered phase emerges for  $T>T_{\mathrm{c}}$ whose  phenomenology is governed by a spectral peak, with  photons tending to occupy states with $\mathpzc{v}_\perp\ll \mathpzc{v}_\parallel$, i.e.  $\tan(\vartheta)\ll 1$.  

To further  understand  the described photon phases,  we   determined  the nontrivial angle $\theta^\prime\in[0,\pi]$  that  makes the function $\tan(\vartheta)=0$.   This indeed occurs if the refraction index in Eq.~\eqref{relatiorn} satisfies 
\begin{equation}\label{criticalindex}
\mathpzc{n}_-^\prime\equiv \mathpzc{n}_-(\omega,\theta^\prime)=\frac{1}{\sqrt{2}v_{\mathrm{s}}}\sqrt{1+\frac{v_{\mathrm{s}}^2}{c^{\prime 2}}-\frac{m^2}{\omega^2}},
\end{equation} with  $\omega\geqslant m\gamma$ and  $\gamma=(1-v_\mathrm{s}^2/c^{\prime2})^{-1/2}$. The angle  $\theta^\prime$ follows  by assuming the existence of a related momentum $\vert\pmb{q}\vert^\prime=\mathpzc{n}_-^\prime\omega$.  After substituting it into   Eq.~\eqref{DE}, we end up with 
\begin{equation}\label{anglerealtootg}
\begin{split}
\sin^2(\theta^\prime)&=1-\frac{v_{\mathrm{s}}^2n_-^{\prime 2}}{\mathfrak{b}^2}\frac{\omega^2}{m^2}\left[1-\frac{1-\frac{m_*^2}{\omega^2}}{c^{\prime 2}v_{\mathrm{s}}^2n_-^{\prime 4}}\right].
\end{split}
\end{equation}The expression above  is meaningful  if  $0\leqslant \sin^2(\theta^\prime)\leqslant 1$, which is   guaranteed for 
\begin{equation}\label{rangeoffrequency}
 \gamma m\leqslant\omega\leqslant \gamma m\sqrt{1+2\mathfrak{b}^2}.
 \end{equation}  We note that this constraint is necessary but not sufficient to ensure that thermal quanta propagate along $\pmb{B}$.  Only if  $\theta\approx \theta^\prime$  can  the condition  $\mathpzc{v}_\perp\approx0$  be  guaranteed, and then the nontrivial component $\mathpzc{v}_\parallel$ behaves as shown in  Fig.~\ref{fig:3}(d):  At $\omega=\gamma m$  [see Eq.~\eqref{rangeoffrequency}], $\mathpzc{v}_\parallel$ reaches a maximum  $\mathpzc{v}_{\mathrm{max}}= c^\prime$ and   as $\omega$ increases  rapidly falls to slow light states, approaching its minimum $\mathpzc{v}_{\mathrm{min}}=v_{\mathrm{s}}\mathfrak{b}/\sqrt{1+2 \mathfrak{b}^2}\ll c^{\prime}$ (horizontal dotted lines). Equipped with this information, we  computed
\begin{equation}\label{Nparallel_ex}
\begin{split}
N_\parallel&=\frac{2\pi V}{c^{\prime 2}}\int_{\nu_{\mathrm{min}}}  
^{\nu_{\mathrm{max}}} \frac{d\nu\;\nu^2}{e^{2\pi \nu/T}-1}\int_0^\pi d\theta\frac{\mathpzc{n}_-^2}{n_o^2}\frac{\sin(\theta)}{\vert\mathpzc{v}_{\pmb{q},-}\vert}
\end{split}
\end{equation}  where $\nu_{\mathrm{min}}=\gamma m/2\pi$ and $\nu_{\mathrm{max}}=\gamma m (1+2\mathfrak{b}^2)^{1/2}/2\pi$. Equation~\eqref{Nparallel_ex} includes the narrow frequency sector around  $\mathpzc{v}_\parallel\sim c^\prime$. However, this contribution proves to be negligible   compared to that of the slow thermal states. 
 
 \begin{figure}
\includegraphics[width=0.21\textwidth]{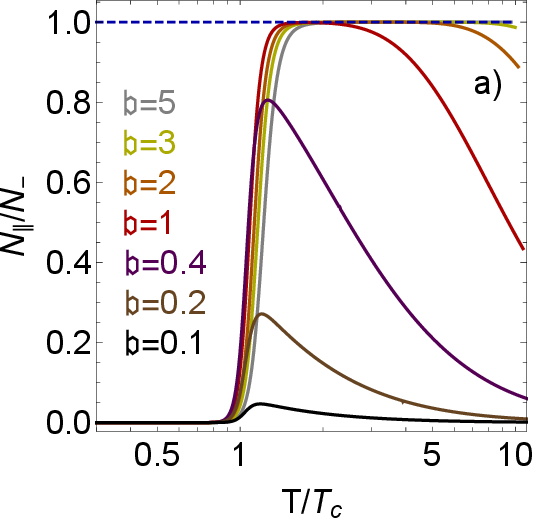}
 \includegraphics[width=0.22\textwidth]{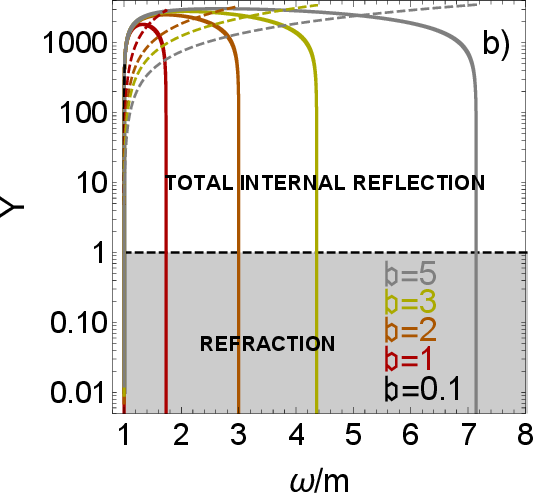}
\caption{ (a) Condensate fraction of extraordinary photons occupying  slow light states as a function of temperature for various $\mathfrak{b}$'s. (b)  Dependence of  Snell's law   with the frequency $\omega$ at $\theta=\theta^\prime$.  For the solid curves, the  label must be understood as $\Upsilon\equiv \mathpzc{n}_-^\prime\sin(\theta^\prime)$, whereas for the dashed ones, it is $\Upsilon\equiv\mathpzc{n}_-^\prime\cos(\theta^\prime)$. } 
\label{fig:4}
\end{figure}
 
The fraction of extraordinary quanta   $N_{\parallel}/N_-$ that occupies this sort of  low-dimensional ($\mathpzc{v}_\perp\approx 0$) macroscopic state  is depicted  in Fig.~\ref{fig:4}(a):   For $T > T_{\mathrm{c}}$, there exist magnetic fields under which the population of slow thermal light occurs abundantly, resembling  a many-body condensate. It is important to note that this condensation differs from the observed Bose-Einstein condensation of photons in dye-filled microcavities \cite{Klaers,Barland,Marelic}  in that it consists of massless excitations occupying a state other than the ground state,  while the latter comprises quanta with an effective mass generated by the cavity cutoff frequency populating the ground state. It  also differs from an early theoretical proposal for slow-light Bose-Einstein condensation coexisting in thermodynamic equilibrium with a Bose-Einstein condensation of an ideal two-level atomic gas \cite{Yurii1,Yurii2,Boichenko}. Taking   $T = 2\; \rm K$ and  $\mathfrak{ b}= 1$, corresponding to $B=2.35\;\rm T$, the density of the condensate is  $ N_\parallel/V\approx 1.9\times 10^8\; \rm photons/cm^3$, whereas the background of ordinary quanta is $N_o/V\approx 10^4\;\rm photons/cm^3$.  The proposed heat flux measurements may disclose the phase transition through their response to  temperature variations  across  $T_{\mathrm{c}}$. 

The existence of the  condensate of slow thermal light raises the question of how it remains trapped within the material. To this end, let us consider  Snell's law. When  applied to the vacuum-DAI interface perpendicular to $\pmb{B}$ it reads  $\mathpzc{n}_{\mathrm{vac}}\sin(\theta_{\mathrm{vac}}^\parallel)=\mathpzc{n}_{-}(\omega,\theta)\sin(\theta)$, where $\mathpzc{n}_{\mathrm{vac}}=1$. Extraordinary photons for which $\mathpzc{v}_\perp\approx0$ are characterized by the refraction index $\mathpzc{n}_-^\prime$ [see Eq.~\eqref{criticalindex}] and the  angle $\theta^\prime$ [see Eq.~\eqref{anglerealtootg}].  The solid curves in  Fig.~\ref{fig:4}b show the behavior of the corresponding  refraction angle  $\sin(\theta_{\mathrm{vac}}^\parallel)=\mathpzc{n}_{-}^\prime(\omega)\sin(\theta^\prime)$ with frequency lying in the interval given in Eq.~\eqref{rangeoffrequency}.  According to our previous conclusion, only minuscule fractions of both fast   ($\mathpzc{v}_\parallel\approx c^\prime$)  and extremely slow    ($\mathpzc{v}_\parallel\approx \mathpzc{v}_{\mathrm{min}}$)   photons can experience refraction and consequently escape into  vacuum. In addition,  Fig.~\ref{fig:4}(b)  provides evidence that the predominant number of slow light states ($ \mathpzc{v}_{\mathrm{min}} <\mathpzc{v}_\parallel\ll  c^\prime$)    become  confined  within the insulator's boundary  due to total internal reflection. The associated  critical angle $\theta_{\mathrm{c}}=\arcsin(1/\mathpzc{n}_-^\prime)$ depends  on $\omega$. At the energy where the  group velocity is minimal, [upper bound in Eq.~\eqref{rangeoffrequency}] $\theta_{\mathrm{c}}=\arcsin(v_{\mathrm{s}}\sqrt{1+2\mathfrak{b}^2}/\mathfrak{b})$. For $\mathfrak{b}=5$, $n_-^\prime\approx 3.5 \times 10^3$ and  $ \theta_{\mathrm{c}}\approx0.3\;\mathrm{mrad}$. Next Snell's law relative to the interface parallel to $\pmb{B}$  can  be expressed as  $\mathpzc{n}_{\mathrm{vac}}\cos(\theta_{\mathrm{vac}}^\perp)=\mathpzc{n}_{-}(\omega,\theta)\cos(\theta)$.  The dashed curves in Fig.~\ref{fig:4}(b)  depict  that  $\mathpzc{n}_{-}^\prime(\omega)\cos(\theta^\prime)>1$ across the entire $\omega$ range, indicating that photons from  the condensate cannot escape to  vacuum via the surface parallel to $\pmb{B}$. This is consistent with the findings shown in the right sector of Fig.~\ref{fig:2}(b).
 
 \section{Conclusions \label{conclusions}}
 
The thermal radiation properties in DAIs are ruled by the strong birefringence that the axion-polariton state transfers to the statistical ensemble. Magnetic field and temperature are suitable parameters for quantum controlling the distribution, strength ,and asymmetry of heat radiation in DAIs. Furthermore, they play a crucial role in manipulating the population of slow, extraordinary photons that could condense into a low-dimensional state of energy transport. The reported thermal features in DAI materials apply to both topological and trivial phases and thus underline their universality. However, we have seen that other thermodynamic quantities, such as free energy and total internal energy, are affected by the $\bar{\theta}$ value [see Eqs.~\eqref{freeenergy} and \eqref{IEnergy}]. Our findings offer prospects for an out-of-contact, directionally tunable mechanism for manipulating heat radiation properties in DAIs.
  
Let us briefly address what to expect from the thermal radiation in the MDCDW phase of dense quark matter. Like DAIs, the Planck radiation law in this phase will be  given by  an expression similar to  Eq.~\eqref{MPRL}.  However, as the axion quasiparticles of this model are massless  \cite{EfrainNP,EfrainEPJC}, the takeoff of the corresponding extraordinary spectrum (see Fig.~\ref{fig:1}c)  is expected to be much more dramatic toward low frequencies, significantly exceeding that of the ordinary mode  and allowing the thermal radiation to be highly asymmetric.  The lack of mass renders it difficult, though,  to foresee whether a peak similarly pronounced as in  DAIs can emerge, along with a corresponding  phase transition and condensation at the infrared.   However, the  anisotropic  heat distribution outlined for DAIs is  expected to be present in  neutron star cores. A thorough investigation of these and other compelling aspects of blackbody radiation in the MDCDW  phase of dense QCD is left for future work.

\begin{acknowledgments}
 
 GQA  gratefully acknowledges the support of the Alexander von Humboldt Foundation. RE acknowledges funding from the Deutsche Forschungsgemeinschaft (German Research Foundation), under Project No.~277101999, TRR 183 (project C01),  and under Germany's Excellence Strategy Cluster of Excellence Matter and Light for  Quantum Computing  EXC 2004/1 Grant No.  390534769.  
 
\end{acknowledgments}

\appendix
\section{Integration domains for  spectral fluxes \label{AppC}}

\begin{figure}
 \includegraphics[width=0.235\textwidth]{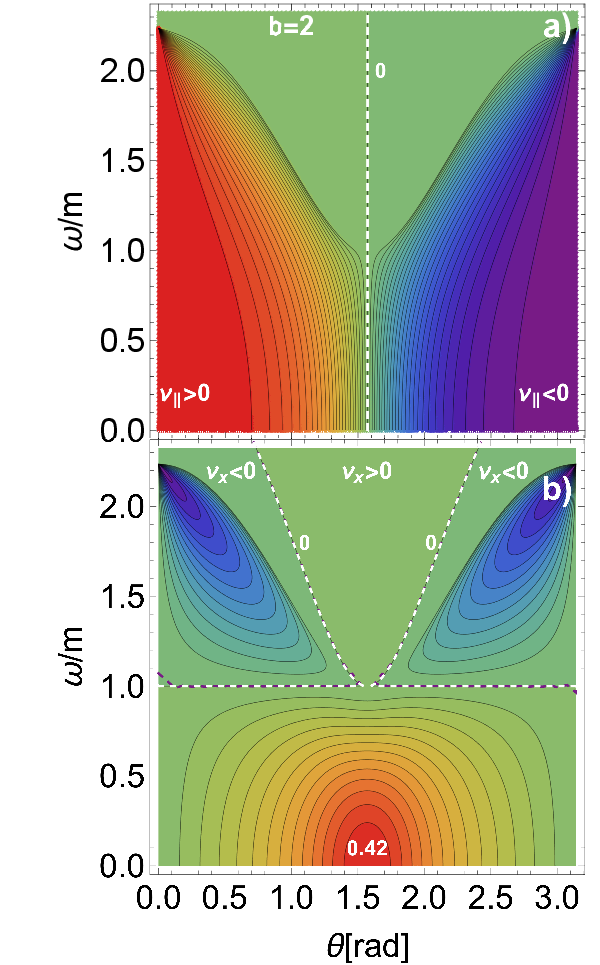}
  \includegraphics[width=0.22\textwidth]{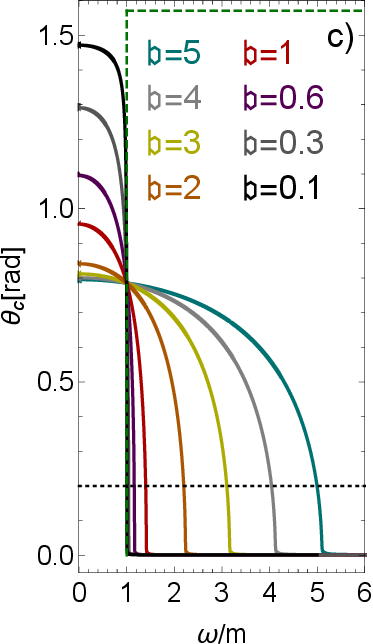}
 \caption{Cartesian components of the extraordinary mode's group velocity  (a)  parallel and (b) perpendicular to the magnetic-field axis as a function of the frequency $\omega$ and the phase angle $\theta$.  (c) Critical angle for total internal reflection of the extraordinary mode parallel  (solid curves) and perpendicular (green dashed curve) to the magnetic field.  The latter is unaffected by the field strength and approaches zero abruptly at  $\omega=m$. The horizontal dotted line represents the critical angle related to ordinary radiation.  \label{fig:5}}
\end{figure}  

Consider  the spectral flux $\mathpzc{s}_{\parallel,-}$ of the extraordinary mode.  Here  $\mathpzc{v}_{\parallel,-}$ is positive when the angular integration is limited to the upper hemisphere, i.e., when $\varphi\in[0,2\pi)$ and $\theta\in [0,\pi/2)$.  Figure~\ref{fig:5}(a) is meant to verify  this statement. It shows how $\mathpzc{v}_{\parallel,-}$ behaves with $\omega$ and $\theta$. The  white dashed line indicates the contour at which $\mathpzc{v}_{\parallel,-}=0$; on its left  $\mathpzc{v}_{\parallel,-}>0$. The situation is different for  $\mathpzc{v}_{x,-}$ which follows from  the first line in  Eq.~\eqref{velocities_components} by inserting a  factor $\cos(\varphi)>0$.  As shown in Fig.~\ref{fig:5}(b),  the remaining angular domain that ensures $\mathpzc{v}_{x,-}>0$ depends on  the value of $\omega$. For $\omega\leqslant m$, the $\theta$ integration can be safety extended over the region $(0,\pi)$.  However, for $\omega\geqslant m $,  the $\theta$ sectors that guarantee  the positive value of $\mathpzc{v}_{x,-}$ is limited by the contour  $\theta^\prime(\omega)$ [see Eq.~\eqref{anglerealtootg}] at which  $\mathpzc{v}_{x,-}=0$ (white dashed curve).  Hence, for $\omega>m$,  $\mathpzc{s}_{\perp,-}=\mathpzc{s}_{x,-}$  has been calculated by considering $\vert\theta\vert <\theta^\prime$. The complementary areas to this sector (covering blue hue) yield $\mathpzc{v}_{x,-}\leqslant 0$, even though $q_x=\vert\pmb{q}\vert\sin(\theta)>0$. The spectral fluxes along the  $\pmb{B}$ direction of  mode $o$ and $+$ can be calculated using the same domain found  for  $\mathpzc{s}_{\parallel, -}$. Conversely,  $\mathpzc{s}_{\perp,o}$ and  $\mathpzc{s}_{\perp,+}$ follow by limiting  $\varphi\in (-\pi/2,\pi/2)$ and $\theta\in (0,\pi)$.

Now  the  spectral flux  $\mathfrak{s}_\parallel^{\mathrm{ref}}=\mathpzc{s}_{\parallel,o}^{\mathrm{ref}}+\mathpzc{s}_{\parallel,-}^{\mathrm{ref}}$ that may refract at the DAI-vacuum interface perpendicular to $\pmb{B}$ is obtained by integrating 
  $\mathpzc{s}_{\parallel,o}$ and $\mathpzc{s}_{\parallel,-}$ over the region where $\varphi\in[0,2\pi)$ and   $\theta\in [0,\theta_{\parallel}]$. Here  $\theta_{\parallel,o}=\arcsin(1/\mathpzc{n}_o)\approx 0.2\;\mathrm{rad}$  is  the critical angle for total internal reflection of the ordinary mode [horizontal dotted line in  Fig.~\ref{fig:5}(c)],  whereas $\theta_{\parallel,-}(\omega)$ is the one of the extraordinary photons. Its dependence on  $\omega$ was  found  numerically by solving the equation $\mathpzc{n}_-(\omega,\theta)\sin(\theta)=1$ [solid curves Fig.~\ref{fig:5}(c)].  When calculating the spectral flux $\mathfrak{s}_\perp^{\mathrm{ref}}=\mathpzc{s}_{x,o}^{\mathrm{ref}}+\mathpzc{s}_{x,-}^{\mathrm{ref}}$  that may refract at the  interface  perpendicular to the $x$ axis, we considered that $\mathpzc{s}_{x,o}^{\mathrm{ref}}=\mathpzc{s}_{\parallel,o}^{\mathrm{ref}}$. The contribution of  extraordinary modes $\mathpzc{s}_{\perp,-}^{\mathrm{ref}}$ depends on the frequency $\omega$  [green dashed curve in Fig.~\ref{fig:5}(c)]. For $\omega<m$,  all quanta can undergo refraction.  As a result, the behavior of the corresponding sector in  Fig.~\ref{fig:2}(b) matches  that of Fig.~\ref{fig:2}(a). For $\omega\geqslant m$, the angular  integration  is limited  to $\varphi\in[\varphi_-,\varphi_+]$ and $\theta\in[\theta_{\perp},\pi-\theta_{\perp}]$, where  
  \begin{equation}
  \varphi_\pm=\pm\arcsin\sqrt{\frac{\cos^2(\theta_{\perp})-\cos^2(\theta)}{\sin^2(\theta)}}.
  \end{equation} 
 The number of photons that can be refracted is constrained  here by total internal reflection, which is characterized  by a critical angle $\theta_{\perp}$ that is  determined  by solving  $\mathpzc{n}_-(\omega,\theta)\cos(\theta)=1$  [green dashed curve in Fig.~\ref{fig:5}(c)].  This procedure resulted in  a $\theta_{\perp}$ very close to $\pi/2$, limiting the  $\theta$ integration to an infinitesimally small region. It prevents extraordinary photons with  $\omega\geqslant m$  from refracting across the surface. Figure~\ref{fig:2}(b) summarizes  the  behavior of  $\mathfrak{s}_\perp^{\mathrm{ref}}$ with $\nu$. It shows an abrupt change in the vicinity of  $\nu=0.48\; \rm THz$ ($\omega=m$).   To the left (right) of this reference value, the spectrum is dominated by extraordinary (ordinary) radiation.

\end{document}